\documentclass[prl,twocolumn,prb]{revtex4}

\usepackage{graphicx}
\usepackage{epsfig}
\usepackage{verbatim}
\usepackage{bm}

\newcommand{\gammabar}{\ensuremath\gamma\kern-0.53em-}

\usepackage{times}
\usepackage{amsmath}
\usepackage{amsfonts}
\usepackage{amssymb}
\usepackage{graphicx}
\usepackage{pst-all}
\usepackage{leftidx}

\usepackage{tikz}

\makeatletter
\def\l@subsubsection#1#2{}
\makeatother

\begin{document}

\title{Superconductivity Induced Topological Phase Transition at the Edge of Even Denominator Fractional Quantum Hall States}
\author{Maissam Barkeshli and Chetan Nayak}
\affiliation{Station Q, Microsoft Research, Santa Barbara, California 93106-6105, USA}
\date{\today}

\begin{abstract}
We show that every even-denominator fractional quantum Hall (FQH) state possesses at least two robust, topologically distinct 
gapless edge phases if charge conservation is broken at the boundary by coupling to a superconductor. 
The new edge phase allows for the possibility of a direct coupling
between electrons and emergent neutral fermions of the FQH state. This can potentially be experimentally
probed through geometric resonances in the tunneling density of states
at the edge, providing a probe of fractionalized, yet electrically neutral, bulk quasiparticles. Other measurable
consequences include a charge $e$ fractional Josephson effect, a charge $e/4q$ quasiparticle blocking 
effect in filling fraction $p/2q$ FQH states, and modified edge electron tunneling exponents. 
\end{abstract}

\maketitle

\it Introduction \rm-- Fractional quantum Hall (FQH) states often
possess a minimal set of robust gapless edge states \cite{wen04,nayak2008}.
The correspondence between the bulk topological properties and the boundary theory is often 
referred to as the bulk-boundary correspondence, and forms a crucial part 
of our understanding of the topological properties of FQH states. 

However, even the minimal boundary phases of FQH states are not generally unique. It has recently been shown that
when the boundary of a FQH state -- or the interface between two FQH states -- contains the same number of left and right movers, 
it is possible under certain conditions for these boundary modes to be fully gapped by backscattering 
in topologically distinct ways \cite{barkeshli2012a,lindner2012,barkeshli2013genon,clarke2013,cheng2012,levin2013,barkeshli2013defect,barkeshli2013defect2}.
More generally, some topologically ordered states can support multiple,
topologically distinct types of gapped boundaries \cite{bravyi1998,bais2009b,beigi2011,kitaev2012,barkeshli2012a,lindner2012,clarke2013,cheng2012,
barkeshli2013genon,wang2012, levin2013,barkeshli2013defect,barkeshli2013defect2,barkeshli2014sledge,lan2015,hung2015} \footnote{See also \cite{kapustin2011,fuchs2012} for mathematical discussions of topological boundary conditions. }.
These topologically distinct gapped boundaries can potentially provide
novel probes of electron fractionalization by allowing direct
conversion of electrons into fractionalized quasiparticles \cite{barkeshli2014sledge}. 
Domain walls between different boundaries are non-Abelian defects hosting exotic zero modes \cite{barkeshli2012a,lindner2012,cheng2012,barkeshli2013genon,fendley2012,clarke2013,barkeshli2013defect,barkeshli2013defect2,barkeshli2014SDG}. 

Even when the boundary theory is gapless and fully chiral, it has been discovered 
that it may be possible to have multiple distinct types of chiral gapless boundaries \cite{cano2014}. 
The $\nu  = 8$ IQH state, for example, has been found to support two edge phases, 
both with chiral central charge $c = 8$, which exhibit different universal tunneling exponents. 
When the stable gapless edge states are not fully chiral, it is well-known that disorder can induce
different (but topologically equivalent) gapless edge phases \cite{kane1994}. 

In this paper, we show that every even-denominator FQH state supports at least two topologically 
distinct gapless edge phases if charge conservation is broken at the boundary by coupling to a superconductor. 
We explicitly demonstrate this using the chiral Luttinger liquid theory of the edge in two examples, the Moore-Read 
Pfaffian FQH state \cite{moore1991} and the two-component $(331)$ state. 
The new boundary phase admits a novel observable phenomenon, which was recently discussed in the context of quantum spin liquids \cite{barkeshli2014sledge}:
a direct quantum mechanical coupling between electrons and topologically non-trivial neutral fermionic excitations of the FQH state, which can be probed
through tunneling measurements. Other measurable consequences include a charge $e/4$ quasiparticle blocking effect in quantum point contacts (QPC),
a charge $e$ fractional Josephson effect, and modified edge electron tunneling exponents. 

\it Moore-Read Pfaffian \rm -- A useful description of the Moore-Read Pfaffian state at filling $\nu = 1/2$ is
in terms of a $p_x + i p_y$ paired state of composite fermions \cite{read2000,jainCF}. The effective field theory can be described by the 
Lagrangian
\begin{align}
\label{MRtheory}
\mathcal{L} = -\frac{2}{4\pi} a \partial a + \frac{1}{2\pi} (\tilde{a}
+ A_E) \partial a + \mathcal{L}_{cf}(\psi, \psi^\dagger, \tilde{a}), 
\end{align}
where $a$ and $\tilde{a}$ are emergent fluctuating $U(1)$ gauge fields, $A_E$ is the physical electromagnetic gauge field, we have used the notation 
$a \partial a \equiv \epsilon_{\mu\nu\lambda} a_\mu \partial_\nu a_\lambda$, $\psi$ is the composite fermion, and
$\mathcal{L}_{cf}(\psi, \psi^\dagger, \tilde{a})$ is the effective Lagrangian for the paired state of composite fermions.
The pairing of the composite fermions breaks the $U(1)$ gauge field $\tilde{a}$ down to a $Z_2$ gauge field. 
Integrating out $a$ gives rise to the conventional presentation
of the composite fermion effective theory \cite{halperin1993}. It is useful to understand 
this field theory as arising from a parton construction in which the electron 
operator is written as $c = b \psi$, where $b$ is a charge $e$ boson, and $\psi$ is the composite fermion. 
This introduces the $U(1)$ gauge field $\tilde{a}$, associated with the 
gauge redundancies $b \rightarrow e^{i\theta} b$, $\psi \rightarrow e^{-i\theta} \psi$. The Moore-Read state corresponds to a mean-field
state of the partons where $b$ forms a $\nu = 1/2$ bosonic Laughlin state, while $\psi$ forms a $p_x + i p_y$ paired superconductor. 
The conserved current $j_\mu^b \equiv \frac{1}{2\pi} \epsilon_{\mu\nu\lambda} \partial_\nu a_\lambda$ corresponds 
to the conserved current of $b$; the first term in (\ref{MRtheory}) can then be interpreted as the effective field theory 
for the $1/2$ Laughlin state of $b$ \cite{wen04}. 

If the charge $e$ boson $b$ condenses, then $c = \langle b \rangle \psi$, so that $c$ and $\psi$ effectively become identified as operators.
The emergent gauge symmetry associated with the gauge field $\tilde{a}$, which was previously broken to $Z_2$, is now broken completely. 
Consequently, the electrons now form a $p_x + i p_y$ superconductor, because 
$\langle c c \rangle \sim \langle \psi \psi \rangle$ \footnote{This relates the quantum phase transition between the Moore-Read Pfaffian state and a $p_x + i p_y$ superconductor to the transition between the bosonic $1/2$ Laughlin state and a Bose superfluid \cite{barkeshli2012hlr}.} .
Therefore, as the charge $e$ boson $b$ in the MR state approaches the boundary with a $p_x + i p_y$ state, 
it will disappear into the $b$ condensate at the boundary, 
\it despite being a topologically non-trivial excitation in the bulk of the Moore-Read state\rm. 
On the other hand, at the boundary of the MR state with vacuum, $b$ remains uncondensed and thus 
cannot disappear at the boundary. Since these two boundaries are distinguished by the condensation of a topologically
non-trivial quasiparticle $b$, these considerations suggest the possibility of two topologically distinct, but 
chiral and gapless, boundary phases for the MR state.

In order to establish the existence of the two topologically distinct chiral boundary phases more rigorously,
let us now consider the edge theory of the MR state directly. The usual edge of the MR Pfaffian state is described by a chiral boson $\phi_L$
and a chiral Majorana fermion $\psi_L$, with Lagrangian
\begin{align}
\mathcal{L}_{\text{edge}} = -i \psi_L (\partial_t + v_1 \partial_x) \psi_L -
\frac{2}{4\pi} \partial_x \phi_L \partial_t \phi_L - v_2 (\partial_x \phi_L)^2 ,
\end{align}
where $v_1$, $v_2 > 0$ set the velocities of the edge modes. 
Quantization of this theory yields the commutation relations
$[\phi_L(x), \phi_L(y)] = i \frac{\pi}{2} \text{sgn}(x-y)$. 
Importantly, this edge theory also possesses a $Z_2$ gauge symmetry, associated with the transformations 
$\psi_L \rightarrow -\psi_L$ and $\phi_L \rightarrow \phi_L + \pi/2$. 
This $Z_2$ gauge symmetry is inherited from the $Z_2$ gauge symmetry of the bulk, 
and ensures that all local operators on the boundary must be obtained by operator products 
of the $Z_2$ gauge-invariant electron operators $\psi_L e^{i 2\phi_L}$. Consequently, we 
denote the conventional edge theory of the Moore-Read Pfaffian as
$[\text{MF}\times U(1)_2] / Z_2$. MF, which denotes the edge theory for a
$p_x + i p_y$ superconductor, describes the single chiral Majorana mode, with chiral central
charge $1/2$. We emphasize that the twist operator of the usual Ising
conformal field theory is not present in the MF theory, as is expected
because the twist operator (vortex) in the $p_x + i p_y$ state is a
confined excitation, and becomes deconfined only after the $Z_2$ gauge
symmetry is implemented. The operator $e^{i 2\phi_L}$ is the representation in the edge theory
of the charge $e$ boson $b$ that was introduced in the parton construction of the previous section. 

We wish to consider a different boundary phase, where the $Z_2$ gauge symmetry is 
spontaneously broken on the boundary. To see this explicitly, 
let us introduce two additional pairs of counterpropagating modes
\begin{align}
\mathcal{L}_{\text{recon}} = & -\frac{1}{4\pi} (\partial_x \varphi_{L\alpha} \partial_t \varphi_{L\alpha} 
- \partial_x \varphi_{R\alpha} \partial_t \varphi_{R\alpha})
\nonumber \\
&- V_{IJ}^{\alpha\beta} \partial_x \varphi_{I\alpha} \partial_x \varphi_{J\beta},
\end{align}
where $I,J = L/R$ indicates the chirality of the modes, $\alpha,\beta = 1,2$, and 
$V_{IJ}^{\alpha\beta} $ parametrizes density-density interactions between the modes. 
Physically, $\mathcal{L}_{\text{recon}} $ can arise from edge reconstruction. 
The electron operators on these reconstructed edge modes are given by $c_{L\alpha} \sim e^{ i \varphi_{L\alpha}}$, 
$c_{R\alpha} \sim e^{ - i \varphi_{R\alpha}}$. The fields satisfy 
$[\varphi_{I\alpha}(x), \varphi_{J\beta}(y)] = \pm \delta_{IJ} \delta_{\alpha\beta}i \pi \text{sgn}(x-y)$. 
There are also possible density-density interactions between $\phi_L$ and $\varphi_{I\alpha}$,
of the form $\partial_x \phi_L \partial_x \phi_{I\alpha}$, which are not explicitly included above. 

We emphasize that at this stage we use these two pairs of modes 
as a theoretical convenience to explicitly demonstrate the conclusions. 
Similar arguments can be made using a single pair of reconstructed
edge modes (see Appendix) or none at all. 

Let us consider the following backscattering terms:
\begin{align}
\label{eqn:cosine-terms}
\mathcal{L}_{\text{back}} = & \cos(\varphi_{R-} + \varphi_{L-})
+ \lambda_1 \cos(\varphi_{R+} + \varphi_{L+} )
\nonumber \\
&+ \lambda_2 \cos(4 \phi_L + 2 \varphi_{R+} ),
\end{align}
where $\varphi_{I, \pm} = \varphi_{I1} \pm \varphi_{I2}$, for $I = L/R$. 
The term $\lambda_2$ carries charge $2e$, and thus can only occur if charge conservation is broken at the boundary
by coupling to a superconductor \footnote{Note that $\phi_L$ does not carry spin. Therefore in the present discussion
$e^{i\varphi_{R+}}$ must be either spin singlet or triplet, depending on whether the superconductor has spin singlet or 
triplet pairing. In the singlet case, this requires $\varphi_{I, \alpha}$ to carry spin $\uparrow$ or $\downarrow$ depending on whether
$\alpha = 1,2$. In the triplet case, $\varphi_{I,1/2}$ are both spin polarized.} . The scaling dimensions of these operators can be tuned by tuning the density-density
interactions. When $\lambda_1$ is more dominant, the reconstructed 
modes $\phi_{I\alpha}$ are fully gapped, and we obtain the usual MR Pfaffian edge theory, $[\text{MF}\times U(1)_2] / Z_2$, as described above. 
However, when $\lambda_2$ is more dominant, $(4 \phi_L + 2 \varphi_{R+} )$ is pinned to a constant value. This is possible
because $(4 \phi_L + 2 \varphi_{R+} )$ commutes with the argument of the first cosine term:
$[4 \phi_L(x) + 2 \varphi_{R+}(x) , 4 \phi_L(y) + 2 \varphi_{R+} (y) ] = 0$, and therefore both cosine terms 
can simultaneously pin their arguments to a constant value. In particular, 
\begin{align}
\langle e^{i( 2 \phi_L + \varphi_{R1} + \varphi_{R2})} \rangle \neq 0 
\end{align}
spontaneously breaking the $Z_2$ gauge symmetry $\phi_L \rightarrow \phi_L + \pi/2$. 

To understand the nature of the resulting edge theory, observe that
$\varphi_{L+}$ is the remaining gapless mode. Projecting out the relative density fluctuations associated with $\varphi_{L-}$
leaves behind a $U(1)_2$ chiral boson mode $\varphi_L = \varphi_{L+}/2$, described by the Lagrangian
$\mathcal{L}_{L} = -\frac{2}{4\pi} \partial_x \varphi_{L} \partial_t \varphi_{L} - v_L (\partial_x \varphi_{L})^2$,
where the charge density is $\rho = \frac{2e}{2\pi} \partial_x \varphi_{L}$, $e^{i2 \varphi_L}$ is a charge $2e$ boson, 
and $e^{i \varphi_L}$ is a charge $e$ semion.
Note that the fields $\varphi_{L\alpha}$ satisfy the periodicity condition
$\varphi_{L\alpha} \equiv \varphi_{L\alpha} + 2\pi$. When $\varphi_{L-}$ is pinned,
$\varphi_{L+}$ has the periodicity condition $\varphi_{L+} \equiv \varphi_{L+} + 4\pi$
but  $\varphi_{L+} \not\equiv \varphi_{L+} + 2\pi$ since that would be a shift of
$\varphi_{L1}$ and $\varphi_{L2}$ by $\pi$. Hence,
$e^{i \varphi_L}=e^{i \varphi_{L+}/2}$ is invariant under the periodicity condition
and, therefore, is an allowed edge excitation.

The resulting edge theory thus contains a chiral Majorana mode and a $\nu=1/2$ bosonic edge field 
that are decoupled from each other, which we label as MF $\times U(1)_2$. Relative to 
the previous edge theory $[ \text{MF} \times U(1)_2]/Z_2$, we see that
the new edge has lost the $Z_2$ gauge symmetry. The two different phases are 
topologically distinct because they are distinguished only by whether the nonlocal operator
$e^{i( 2 \phi_L + \varphi_{R+})}$ acquires an expectation value. Since all local operators are $Z_2$ 
gauge invariant, it follows that no local operator can distinguish the two boundary phases.

Note that in the presence of superconductivity, we can also add the charge $2e$ operator 
$\cos(\varphi_{R+})$ to the edge theory. Since $\varphi_{R+}$ is a chiral field, such a term cannot gap
out any modes, but can give an expectation value to the bosonic operator $e^{i \varphi_{R+}}$. 
Now, since $\langle e^{2i \phi_{L} + i \varphi_{R+}} \rangle \neq 0 $ in the new edge phase,
and $\langle e^{i \varphi_{R+}}\rangle \neq 0$, it follows that we should have 
$\langle e^{i 2\phi_L} \rangle \neq 0$. 

Let us consider the fate in the new MF $\times U(1)_2$ edge theory
of the charge $e/4$ non-Abelian quasiparticle, which can be represented
using the operator $\sigma e^{i \phi_L/2}$. Since 
$\phi_L$ does not commute with the argument of the $\lambda_2$ cosine in Eq. (\ref{eqn:cosine-terms})
which is dominant in this phase, $\sigma e^{i \phi_L/2}$ creates
a gapped excitation and, therefore, has exponentially-decaying correlations. In other words,
the charge $e/4$ quasiparticle cannot be added to the edge at low energies. This does not violate
Laughlin's flux insertion argument because $U(1)$ charge conservation is broken on the boundary through superconductivity.
(The only quasiparticle creation operators involving $\phi_L$ that commute with
the argument of the $\lambda_2$ cosine in Eq. (\ref{eqn:cosine-terms}) are of the form
$e^{in(2\phi_L + \varphi_{R+})}$, and these operators take a constant value in the edge phase
dominated by $\lambda_2$.)

\it (331) state \rm-- The $(331)$ state \cite{wen04} also possesses two topologically distinct fully chiral boundary phases. 
Similar to the MR state, the $(331)$ state can be understood in terms of composite fermions by attaching $2$ units
of flux quanta to two flavors of electrons, leading to two flavors of composite fermions, $\psi_\alpha$, where 
$\alpha = \uparrow, \downarrow$ is a flavor (e.g. spin or layer) index. The $(331)$ state is a state
where $\psi_\alpha$ both form a topologically non-trivial $p_x + i p_y$ paired state \cite{read2000}. The effective field theory is 
identical to that of the Moore-Read Pfaffian, except that the composite fermion sector contains two flavors. Therefore, just as for 
the MR Pfaffian, the two distinct boundary phases can be understood in terms of whether the $Z_2$ gauge symmetry is broken
on the edge by the condensation of the charge $e$ boson. 

To provide a more precise account, let us consider the edge theory of the $(331)$ state. The composite fermion description 
motivates a description of the edge theory that is similar to the MR Pfaffian, but now in terms of two chiral Majorana 
fields, $\psi_{L1}$ and $\psi_{L2}$, and a chiral $U(1)_2$ bosonic field $\phi_c$. The two chiral Majorana fields can be bosonized
$\psi_{L1} + i \psi_{L2} \sim e^{i \phi_n}$, where $\phi_n$ is a neutral chiral boson field. The $Z_2$ gauge symmetry now
is associated with the transformation $\phi_n \rightarrow \phi_n + \pi$, $\phi_c \rightarrow \phi_c + \pi/2$, and
ensures that all physical operators can be written in terms of operator products of the gauge-invariant 
electron operators, $e^{\pm i \phi_n} e^{i 2 \phi_c}$. Therefore, the edge theory can be understood as $[U(1)_2 \times U(1)_1]/Z_2$. 
We will show below that there is another edge phase where the $Z_2$ gauge symmetry is broken, and the edge theory
changes to $U(1)_2 \times U(1)_1$. 

In order to show this, we use the more conventional description of the edge theory of the (331) state, which is in terms of 
two chiral boson fields, $\tilde{\phi}_{L1}$ and $\tilde{\phi}_{L2}$, together with a $K$-matrix 
$K = \left(\begin{matrix} 3 & 1 \\ 1 & 3 \end{matrix} \right)$ \cite{wen04}:
\begin{align}
\mathcal{L}_{331} = -\frac{K_{IJ}}{4\pi} \partial_x \tilde{\phi}_{LI} \partial_t \tilde{\phi}_{LJ}  - V_{IJ} \partial_x \tilde{\phi}_{LI} \partial_x \tilde{\phi}_{LJ}
\end{align}
These fields satisfy the commutation relations $[\tilde{\phi}_{LI}(x), \tilde{\phi}_{LJ}(y)] = i \pi K^{-1}_{IJ} \text{sgn}(x-y)$. 
The two electron operators are given by $c_I \sim e^{i K_{IJ} \tilde{\phi}_J}$.

As before, we can consider two pairs of counterpropagating edge modes, described again by the Lagrangian $\mathcal{L}_{\text{recon}}$, 
and we can consider the backscattering terms 
\begin{align}
\mathcal{L}_{\text{back}} = &\cos( \varphi_{R-} + \varphi_{L-}) + \lambda_1 \cos(\varphi_{L+} + \varphi_{R+})
\nonumber \\
&+ \lambda_2 \cos(4 \tilde{\phi}_{L+} + 2 \varphi_{R+}).
\end{align}
Obseve that the $\lambda_2$ term carries charge $2e$ and can only occur in the presence of superconductivity at the boundary of the FQH state. 
When $\lambda_1$ is dominant, the reconstructed edge modes are gapped and we have the usual edge phase of the $(331)$ state. 
When $\lambda_2$ is dominant, the arguments of the cosine terms in $\mathcal{L}_2$ are pinned to a constant value in space.
This, in particular, implies $\langle e^{i 2 (\tilde{\phi}_{L1} + \tilde{\phi}_{L2}) + i(\phi_{R1} + \phi_{R2})}\rangle \neq 0$, which physically 
corresponds to the condensation of the charge $e$ boson of the $(331)$ state, together with two electron operators from the reconstructed edge modes.

There are two remaining gapless modes: $\tilde{\phi}_{L-}$, $\varphi_{L+}$. These two terms commute with each other;
$[\tilde{\phi}_{L-}(x), \tilde{\phi}_{L-}(y)] = i \pi sgn(x-y)$, implying that $\tilde{\phi}_{L-}$ is a $U(1)_1$ chiral boson mode. 
The fate of $\varphi_{L+}$ is identical to the case of the MR state, and is described by a chiral $U(1)_2$ boson mode. Thus, 
we see that the resulting edge theory is $U(1)_2 \times U(1)_1$, indicating that the $Z_2$ gauge symmetry of the original 
$[U(1)_2 \times U(1)_1]/Z_2$ has been broken. 

As in the case of the MR state, one can verify that here also the charge $e/4$ quasiparticle, described now by the operator 
$e^{i \tilde{\phi}_{LI}}$, is a gapped excitation in the new boundary phase. 

\begin{figure}
	\centering
	\includegraphics[width=3.3in]{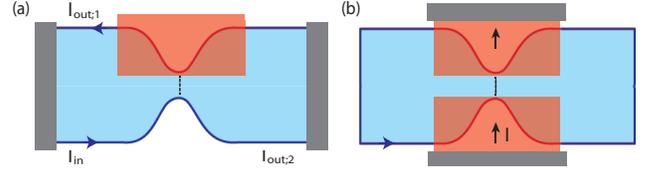}
	\caption{\label{qpc} (a) Experimental setup for probing quasiparticle backscattering across a QPC. Blue edges indicate the original edge phase and red edges
indicate the new, superconductivity-induced edge phase. Dashed lines indicate quasiparticle tunneling. 
The light red block is a superconductor. (b) Setup for measuring fractional Josephson effect in 
S-FQH-S junction.
}
\end{figure}

\it Charge $e/4$ quasiparticle blocking \rm -- Both the MR state and the $(331)$ state possess a charge $e/4$
quasiparticle. In the $Z_2$ gauge symmetry breaking edge state, this charge $e/4$ quasiparticle is gapped and does not exist
at low energies. This effect can be measured through standard quantum point contact experiments, which measure fractional
charge through shot noise \cite{kane1994,picciotto1997}. Specifically, one can consider the setup shown 
in Fig. \ref{qpc}. If both of the outer edges are in the conventional edge state, with the $Z_2$ gauge symmetry 
intact, then shot noise measurements would indicate a minimal quasiparticle charge of $e/4$.
However, if either of the outer edges is in the $Z_2$ gauge symmetry breaking phase, then $e/4$ quasiparticles cannot tunnel across
the QPC at low energies. Interestingly, as we show in the Appendix, the $e/2$ quasiparticles also cannot tunnel across the QPC at low energies;
instead, the minimal charge that can tunnel across the QPC carries charge $e$. Therefore, one expects a crossover from 
charge $e$ to charge $e/4$ quasiparticle tunneling as the voltage across the QPC is increased above the gap to the $e/4$ quasiparticles. 

\it Direct coupling to emergent neutral fermions \rm -- As described above, the $(331)$ and MR Pfaffian states both possess a neutral 
fermionic excitation, which can be interpreted as the Bogoliubov quasiparticle of the composite fermion paired state. 
A profound consequence of the existence of the new chiral boundary phase
is that electrons can directly couple to these neutral fermions at the boundary. This implies that as an electron is injected into the
boundary at energies above the gap of the FQH state, it can leave its charge at the boundary and coherently propagate into the bulk
directly as the neutral fermion. 

To see this more precisely in the context of the edge theory calculations considered in the previous sections,
let us consider the low-energy effective Lagrangian for electron tunneling from a superconductor into the edge of the MR state:
\begin{align}
\mathcal{L}_{\text{tun}} = t_1 c_{\text{sc}}^\dagger \psi_L e^{i 2\phi_L}  + t_2 c_{\text{sc}}^\dagger \psi_L e^{i 2 \phi_L}
\Delta^* e^{i \varphi_{R+}} + H.c.,
\end{align}
where $c_{\text{sc}}$ is the electron annihilation operator in the superconductor and
$\Delta$ is the Cooper pair creation operator in the superconductor. 
The first term is just a single electron tunneling process from the superconductor to the edge of the FQH state. The second term is a 
multi-electron tunneling process, where the electron tunnels from the superconductor into the MR state, while a Cooper pair tunnels
from the reconstructed edge modes into the superconducting condensate. Recall that in the new edge phase, 
$\langle e^{i 2 \phi_L + i \varphi_{R+}} \rangle \neq 0$. 
Therefore, in the new boundary phase we can replace it by a constant, and $\mathcal{L}_{\text{tun}}$ becomes
\begin{align}
\mathcal{L}_{\text{tun}} \rightarrow  t_1 c_{\text{sc}}^\dagger \psi_L e^{i 2\phi_L}  + t_2 \Delta^* \langle e^{i 2 \phi_L + i \varphi_{R+}} \rangle c_{\text{sc}}^\dagger \psi_L + H.c. 
\end{align}
This demonstrates a direct coupling between the electron and the neutral fermion of the FQH state. If the edge theory
also contains the electron pairing term $\cos(\varphi_{R+})$ by itself, then $\langle e^{2i \phi_L}\rangle \neq 0$, so that 
\begin{align}
\label{coupling}
\mathcal{L}_{\text{tun}} = t_{\rm eff}\, c_{\text{sc}}^\dagger \psi_L + H.c. 
\end{align}
where $t_{\rm eff} = t_1 \langle e^{2i \phi_L}\rangle + t_2 \Delta^* \langle e^{i 2 \phi_L + i \varphi_{R+}}\rangle$.
Such a direct coupling between electrons and the emergent neutral excitations has potential experimental consequences, two of which
we describe below. For the sake of concreteness, we assume below that the superconducting
gap is smaller than the bulk FQH gap. 

If electrons from the superconductor tunnel into the edge at energies that are small compared with the bulk FQH gap, then according to 
(\ref{coupling}) they can tunnel directly into the neutral fermion mode $\psi$. Since this has a scaling dimension $\Delta_\psi = 1/2$, 
the exponent for electron tunneling into the edge will differ from that of the usual edge theory, where the electron operator has a scaling dimension 
$3/2$. In particular, this implies that the tunneling current $I_{\text{tun}}$ from the superconductor at a point contact will
will scale at low voltages $V$ like $I \propto V$ in the new edge phase, as compared with $I \propto V^3$ in the conventional edge phase. 

\begin{figure}
	\centering
	\includegraphics[width=2.2in]{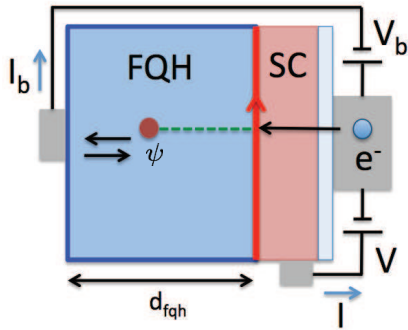}
	\caption{\label{tomasch} Possible experimental setup for detecting coherent transmutation of electrons into neutral emergent fermions of the FQH state. 
Red edge indicates new superconductivity-induced edge phase. Dashed green line indicates the $Z_2$ gauge field that couples to the emergent fermion in the bulk
of the FQH state, and which terminates at the new boundary phase. 
}
\end{figure}

If an electron from the superconductor tunnels into the edge at energies that are larger than the bulk FQH gap, then (\ref{coupling}) allows for the possibility
that the electron enters the bulk of the FQH state as the neutral fermion, coherently leaving its charge behind at the boundary. This basic phenomenon was described
recently in the context of quantum spin liquids in \cite{barkeshli2014sledge}, as a way of detecting fractionalization in both gapped and gapless spin liquids. 
The same basic experiment proposed in \cite{barkeshli2014sledge} can be adapted to the present context. 
Specifically, we can consider the setup shown in Fig. \ref{tomasch}.
In this figure, we show two different mesoscopic electrical transport measurements
that can be performed on a superconductor in contact with a MR FQH state.

In the first, the electrical current $I(V)$ through an NS junction is measured at voltages $V$ greater than
the superconducting gap $\Delta$. In the absence of the FQH state, this would simply be a measurement of
the tunneling density-of-states (TDOS) in the superconductor. However, the current $I(V)$ is modified
by the presence of a FQH state in contact with the superconductor (although neither lead is directly connected
to the FQH state in this measurement), due to Friedel oscillations of neutral $\psi$ fermions.
When an electron enters the superconductor, there is some amplitude for it to tunnel
coherently into the FQH state, where it becomes a neutral $\psi$ fermion, shedding its electrical charge at the edge.
This neutral fermion can be reflected by the left edge of the FQH state, setting up standing waves
at a wavevector larger than the neutral fermion Fermi wavevector $k_F$ by an amount proportional to
the difference between the voltage and the neutral fermion energy gap. As the voltage is increased,
the wavevector of these standing waves increases and the amplitude for
the neutral fermion to return to the right edge of the FQH state oscillates,
with period proportional to $\Delta V \propto 1/d_{\rm fqh}$. This oscillatory component sits
on top of a larger background component that includes processes in which the current flows directly in the superconductor
from one contact to the other without passing through the FQH state. This background also includes processes
in which the $\psi$ excites edge quasiparticles (at the edge between the SC and the FQH state)
while tunneling into the FQH state. As a result of the spread of momenta that can be carried away by 
the excited edge quasiparticles, this background does not exhibit oscillations. If the noise in this background contribution
is smaller than the oscillatory signal that is enabled by the direct coupling between the electron and the neutral fermion,
then this measurement would indicate a direct coupling between the electron in the superconductor and a stable excitation in the bulk of the
FQH state. In order to demonstrate that this stable excitation is not an electron, but rather a topologically
non-trivial, fractionalized excitation, one also requires the following second measurement. 

In the second measurement, the electrical current ${I_b}(V)$ is measured through an N-S-FQH junction,
as shown in Fig. \ref{tomasch}: one normal lead is connected to the superconductor while the other is connected to the FQH state. In order for
current to pass through the FQH state, a $\psi$ particle must excite a quasiparticle at the left edge
of the FQH state in order to form an electron and enter the left lead. This washes out the oscillatory
dependence on $V$, in a manner analogous to the second type of background process mentioned in the 
previous paragraph. The absence of oscillations in $I_b(V)$ would rule out the possibility that oscillations in 
$I(V)$ are due to electronic excitations, and would thus imply the existence of the fractionalized neutral fermion 
excitation in the FQH state. 

We note that the fact that the topologically non-trivial, neutral fermion in the
bulk of the topological state becomes identified with the electron
operator at the boundary of the phase can be seen to give a physical
interpretation to some recent mathematical developments regarding the
notion of fermion condensation in topological quantum field theories
\cite{walker2015}. 

\it Charge $e$ fractional Josephson effect \rm-- Let us consider a superconductor - FQH - superconductor junction, as shown in Fig. \ref{qpc}b, where
the superconductivity induces the new $Z_2$ gauge-symmetry breaking edge phase. Since this edge phase is
characterized by the condensation of charge $e$ bosons, they can coherently tunnel
between the superconducting condensates on either side, inducing a 
charge $e$ fractional Josephson effect. Specifically, let us consider the case of the MR edge
state, with a voltage $V$ applied between the superconductors in Fig. \ref{qpc}b, and 
consider the following tunneling term between the outer edges, across the bulk FQH state:
\begin{align}
\delta \mathcal{L}_{\text{tun}} = \Gamma e^{i e V t} e^{i 2\phi_L + \varphi_{R-}} e^{i 2 \phi_R + \varphi_{L-}} + H.c.
\end{align}
Here, $\phi_L, \varphi_{R\pm}$ are as defined previously, describing modes on the upper edge, while $\phi_R$, $\varphi_{L\pm}$ 
describe similar edge modes on the lower edge. This term tunnels a charge $e$ boson, together with a neutral boson 
from the reconstructed edge modes $\varphi_{R/L, \pm}$, across the bulk FQH state. The tunneling amplitude 
$\Gamma \propto e^{-L/\xi}$, where $L$ is the width of the FQH state and $\xi$ is a correlation length set by the gap of the 
charge $e$ boson in the bulk. This term induces an AC Josephson effect with a frequency $\omega_J = e V$, which is half the usual
Josephson frequency in superconductor - insulator - superconductor junctions. A similar charge $e$ Josephson effect 
was proposed previously as a probe of charge fractionalization in the insulating state of the high Tc cuprates \cite{senthil2001}. 

\it Edge topological phase transition \rm -- Since the two edge phases discussed above are topologically distinct, there must 
be a quantum phase transition between them. Since the transition involves a discrete $Z_2$ gauge symmetry breaking, we expect that
it should map onto a 1+1 dimensional Ising phase transition. The edge central charge at the transition is then expected to be
$(c_L, c_R) = (2,1/2)$ and $(5/2, 1/2)$ for the case of the Moore-Read and (331) states, respectively. Here, $c_{L/R}$ are the left and right
central charges of the edge theory. 

\it Generalizations to other even-denominator FQH states \rm -- The above discussion readily generalizes
to every even-denominator FQH state, at filling fraction $\nu = p/2q$, where $p$ and $q$ are integers and $p$, $2q$ are coprime. Laughlin's flux insertion argument implies that
such a state necessarily possesses a quasiparticle excitation with odd integer charge $p$ and bosonic statistics, associated with the insertion
of $2q$ units of flux. By binding with $p-1$ electrons, we can obtain a charge $e$ bosonic quasiparticle $b$. Binding one additional electron then 
also implies the existence of an electrically neutral fermionic excitation $\psi$.  In terms of these fractionalized degrees of freedom, the electron
operator is written as $c = b \psi$. Such a system necessarily has at least an emergent $Z_2$ gauge symmetry in the low energy effective field theory, 
associated with the transformations $b \rightarrow -b$, $\psi \rightarrow - \psi$, which keep the electron operator invariant. 

If charge conservation is broken at the boundary by coupling to a superconductor, the boundary can support two topologically distinct phases,
characterized by whether the boson $b$ has condensed on the boundary. This corresponds to whether the emergent $Z_2$ gauge symmetry is
preserved or broken on the boundary. In general, this boundary will allow the phenomena discussed above: a direct quantum mechanical coupling between
electrons and the emergent neutral fermions of the FQH state, a charge $e$ fractional Josephson effect, and a charge $e/4q$ quasiparticle blocking effect. 

{\it Generalization to RR, anti-RR states, and BS states}\rm -- 
There is also a generalization to other non-Abelian fractional quantum Hall states. Consider the
Read-Rezayi (RR) states \cite{read1999} at $\nu=k/(k+2)$ and the anti-Read-Rezayi states \cite{Bishara08b}
at $\nu=2/(k+2)$ for $k$ even, which support a charge-$e$ bosonic excitation. The preceeding analysis
can be adapted concretely to the case of the anti-RR states with $k\equiv 2 \pmod{4}$. The charge sector
is a chiral boson $\phi_L$ at level $(k+2)/2$ and the neutral sector is right-moving $SU(2)_k$.
The electron annihilation operator is $\Phi_{j=\frac{k}{2}} e^{i \frac{k+2}{2}\phi_L}$ and
edge quasiparticle operators are $\Phi_j e^{i(j+N)\phi_L}$, where $\Phi_j$ is the
$SU(2)_k$ spin-$j$ primary field. The creation operator for a charge-$e$ boson is simply
$e^{i \frac{k+2}{2}\phi_L}$ (assuming $k\equiv 2 \pmod{4}$), where $\phi_L$ is the charged
boson field in the edge effective field theory.
Following a construction that is nearly identical to
that explained around Eq. (\ref{eqn:cosine-terms}) and in the Appendix,
we see that there is an edge phase in which this charge-$e$ boson condenses at the edge.
In the resulting edge phase, all of the quasiparticle operators with $j \in \mathbb{Z} + \frac{1}{2}$
are confined. Only integer spin fields $\Phi_j e^{i(j+N)\phi_L}$ remain deconfined.
We believe that a similar phase can occur at the edge of the RR state, but we do not, at present, know
how to analyze this case because the charge-$e$ boson annhilation operator is a product of a chiral boson vertex
operator and a $j=k/2$, $m=1$ parafermion field. A similar issue technical difficulty arises in the case of
the RR and anti-RR states for $k\equiv 0 \pmod{4}$. One exception is the $k=4$ RR state, which
has a representation as $[U(1)\times U(1)]/\mathbb{Z}_2$ \cite{barkeshli2010};
in the new edge phase, the $\mathbb{Z}_2$ gauge symmetry is broken. Since the edge theory
can be written in terms of chiral bosons, this particular RR state can be analyzed by the methods of this paper.
The Bonderson-Slingerland hierarchy states \cite{bonderson2008}, which are candidates to explain some FQH 
plateaus in the second Landau level in GaAs systems, also all possess a charge $e$ boson and thus fall 
under the considerations of this paper. 

\it Ising topological phase \rm -- Another important example of the considerations of this paper is the Ising topological phase, which can be realized
in the Kitaev honeycomb spin model \cite{kitaev2006}. This phase can be understood in terms of an 
emergent $Z_2$ gauge field coupled to emergent fermions $\psi$ forming a $p_x + i p_y$ 
topological superconducting state.  The boundary is described by the chiral Ising CFT, which can alternatively be thought of as a
chiral Majorana mode charged under the $Z_2$ gauge field. If the Ising phase is realized from an electronic Mott insulator, then 
there also exists a charge $e$ boson $b^\dagger = c^\dagger \psi$, which is also charged under the $Z_2$ gauge field.
The condensation of the charge $e$ boson can drive the boundary into a new edge phase which breaks the $Z_2$ gauge symmetry,
with analogous phenomena to those described above. 

\it Discussion\rm -- The examples discussed in this paper are special cases of a more general phenomenon, where topologically non-trivial bosons 
in a topologically ordered phase can condense, either in the bulk or just on the boundary, to induce bulk or boundary phase transitions. General algebraic
discussions of boson condensation in the bulk of topological phases from the point of view of modular tensor category theory 
(which describes topological phases where the microscopic constituents are bosons)
have appeared in the literature \cite{bais2009,kong2014}. Remarkably, as we have shown, 
similar considerations can be explicitly demonstrated within the chiral edge field theories of a wide variety of experimentally 
observed fermionic FQH states, and imply a host of potentially measurable consequences. 

In the case of the $\nu = 8$ state, it was shown recently that the domain wall between the different gapless edge phases hosts a chirality-protected Majorana zero mode \cite{cano2015}. A similar phenomenon should occur
in the present context, since a charge $e/4$ quasiparticle in the $[\text{MF}\times U(1)_2] / Z_2$ edge
phase, incident on a domain wall to the $\text{MF}\times U(1)_2$ edge phase, cannot be reflected (since the
edge is chiral) and cannot pass into the latter phase, which does not have $e/4$ quasiparticles. Hence, there
must be a zero mode at the domain wall that is capable of absorbing such quasiparticles.
There may be interesting additional phenomena since the zero mode must absorb a non-Abelian
quasiparticle. 

\it Acknowledgments \rm-- We thank Steven Kivelson and Erez Berg for a previous, related collaboration, and for early discussions on this subject. We also thank
Parsa Bonderson and Meng Cheng for comments on the manuscript. 


\appendix

\section{Appendix A: $U(1)_2 \times$ MF with a Single Pair of Reconstructed Edge Modes}
\label{sec:one-reconstruction}

In the main text, we described a route to the new edge phase $U(1)_2 \times$ MF that
involved two pairs of reconstructed edge modes. Here, we show how the phase transition to
this edge phase can also occur with a single pair of reconstructed edge modes. With a single pair of
new $\nu=\pm 1$ edge modes $\varphi_{R,L}$, Eq. (\ref{eqn:cosine-terms}) becomes:
\begin{equation}
\mathcal{L}_{\text{back}} = 
{\tilde \lambda}_1 \cos(\varphi_{R} + \varphi_{L} )
+ {\tilde \lambda}_2 \cos(4 \phi_L + \varphi_{L} - 3\varphi_{R}),
\end{equation}
When ${\tilde \lambda}_1$ dominates, the system is in the conventional edge phase of the MR state.
However, when ${\tilde \lambda}_2$ dominates, the only fields that remain gapless at the edge
are of the form $\Phi_{\rm Ising} \exp{i({n_1} \phi_L + {n_2}\varphi_{L} + {n_3}\varphi_{R})}$ with
$2{n_1} + {n_2} + 3{n_3}=0$ and $\Phi_{\rm Ising}$ an Ising field $1, \sigma$, or $\psi$.
This is an alternative representation of the $U(1)_2 \times$ MF
edge phase. In the main text, we presented a construction with two pairs of reconstructed edge modes
because the remaining gapless edge mode is simply $\varphi_{L+}$.

\subsection{Appendix B: Charge $e/2$ quasiparticle blocking}

In the main text we showed that at the new $Z_2$ gauge symmetry breaking edge, charge $e/4$ quasiparticles cannot tunnel across the bulk of the FQH state from one edge to another, as they are gapped in the new edge phase. It was also mentioned that charge $e/2$ quasiparticles also cannot tunnel from one edge to another across a QPC, but that charge $e$ quasiparticles can. Below we elaborate more on this point in the case of the MR Pfaffian state. The (331) case is analogous and will not be discussed explicitly. 

First, let us consider a QPC geometry, as shown in Fig. 1(a) of the main text. The modes of the upper edge are $\phi_L$, $\varphi_{L/R, \pm}$, as introduced in the main text. For the lower edge it suffices to include just the original modes, which include the $U(1)_2$ charged field $\phi_R$, and the Majorana fermion $\psi_R$. As we discussed, if the upper edge is in the new edge phase, the following operator has acquired a non-zero expectation value:
\begin{align}
\langle e^{i2\phi_L + \varphi_{R+}} \rangle \neq 0. 
\end{align}
The semion in the new edge phase is described by the operator $e^{i \varphi_{L+}/2}$. 

Let us consider quasiparticle tunneling across the QPC. The operator $e^{i \phi_R} e^{i \varphi_{L+}/2}$, which tunnels a charge $e/2$ semion from the bottom edge to the top edge is not allowed because it carries charge $e/2$ (modulo $e$). The operator $e^{i\phi_L +i \varphi_{R+}/2} e^{i\phi_R + i \varphi_{L+}/2}$ conserves charge and is allowed, however it is a nonlocal operator. 
The local operator that tunnels the minimal charge across the QPC tunnels charge $e$ and is described by the operator
$e^{i 2 \phi_L + i \varphi_{R+}} e^{i 2\phi_R +i \varphi_{L+}} \sim \langle e^{i 2 \phi_L + i \varphi_{R+}} \rangle e^{i 2\phi_R +i \varphi_{L+}}$. Including this process gives rise to a term in the Lagrangian:
\begin{align}
\mathcal{L}_{qpc} \sim \lambda \delta(x) \cos(2 \phi_R + \varphi_{L+}) ,
\end{align}
where we have assumed the location of the QPC is at $x = 0$. When this term is relevant, the FQH fluid is cut into two pieces, and now the edge consists of a left and right piece, each of which contain a domain wall between the new edge phase and the old edge phase. Importantly, one has
\begin{align}
\langle e^{i \phi_R + \varphi_{L+}/2} (x = 0) \rangle \neq 0,
\end{align}
which implies that the charge $e/2$ semion $e^{i\phi_R}$ can propagate across the domain wall and continue as the charge $e$ semion $e^{i\varphi_{L+}/2}$. 

\bibliography{TI}

\end{document}